\documentclass[preprint,floatfix,superscriptaddress]{revtex4}
\usepackage{graphicx}
\usepackage{dcolumn}
\usepackage{amsmath}
\usepackage{bm}
\usepackage{color}
\usepackage[titletoc,toc,title]{appendix}

\begin{document}

\title{Stochastic Interactions of Two Brownian Hard Spheres in the Presence of Depletants}

\author{Mehdi Karzar-Jeddi}
\affiliation{Department of Mechanical Engineering, University of
Connecticut, Storrs, CT 06269-3139, USA}

\author{Remco Tuinier}
\affiliation{Van $'$t Hoff Laboratory for Physical and Colloid
Chemistry, Debye Institute, Department of Chemistry, Utrecht University, Utrecht, the
Netherlands}

\author{Takashi Taniguchi}
\affiliation{Graduate School of Engineering, Kyoto University
Katsura Campus, Nishikyo-ku, Kyoto 615-8510, Japan}

\author{Tai-Hsi Fan\footnote{Corresponding author. E-mail:
thfan@engr.uconn.edu}$^,$}
\affiliation{Department of Mechanical
Engineering, University of Connecticut, Storrs, CT 06269-3139, USA}

\begin{abstract}
\noindent A quantitative analysis is presented for the stochastic interactions of a pair
of Brownian hard spheres in non-adsorbing polymer solutions. The hard spheres are
hypothetically trapped by optical tweezers and allowed for random motion near the trapped
positions. The investigation focuses on the long-time correlated Brownian motion. The
mobility tensor altered by the polymer depletion effect is computed by the boundary
integral method, and the corresponding random displacement is determined by the
fluctuation-dissipation theorem. From our computations it follows that the presence of
depletion layers around the hard spheres has a significant effect on the hydrodynamic
interactions and particle dynamics as compared to pure solvent and pure polymer solution
(no depletion) cases. The probability distribution functions of random walks of the two
interacting hard spheres that are trapped clearly shifts due to the polymer depletion
effect.
 The results show that the reduction of the viscosity in the depletion layers around the
 spheres and the entropic force due to the overlapping of depletion zones have a significant
 influence on the correlated Brownian interactions.
\end{abstract}

\date{\today}

\maketitle
\section{Introduction}
The interactions between dispersed colloidal particles in solutions containing
non-adsorbing polymer chains play an essential role in many phenomena and processes
including macromolecular crowding, protein crystallization, food processing, and co- and
self-assembly \cite{zimmerman1993macromolecular, ellis2003join,
doublier2000protein,tanaka2002protein, rossi2011cubic}. Adding non-adsorbing polymer
chains to a dispersion of colloids effectively induces an attractive potential between
the colloidal particles
\cite{asakura1954interaction,asakura1958interaction,vrij1976polymers} and alters their
phase behavior \cite{gast1983polymer,lekkerkerker1992phase,lekkerkerker1994phase,
poon2002physics,koda2002test,fleer2008analytical} and transport properties
\cite{tuinier2006depletion}. These changes originate from the presence of polymer
depletion zones around the colloidal particles. To avoid the reduction of conformation
entropy, polymers prefer to stay away from the colloidal surfaces. Hence depletion zones
appear around the colloidal particles. Within the depletion zone the polymer
concentration is reduced significantly compared to the bulk polymer concentration. The
overlap of depletion layers causes an unbalanced osmotic pressure distribution by the
polymers around the colloids, first understood by Asakura and Oosawa
\cite{asakura1954interaction, asakura1958interaction}. The resulting attractive
potential's range and depth can be tuned by polymer size, concentration, and solution
conditions. In the last few decades many studies on polymer depletion were focused on the
equilibrium aspects of colloid-polymer mixtures, primarily on the depletion interaction,
phase behaviors, microstructure of the colloid-polymer suspensions, and scattering
properties~\cite{gast1983polymer, lekkerkerker1992phase, verma1998entropic,
fuchs2002structure, poon2002physics, koda2002test, dzubiella2002phase, mutch2007colloid,
kleshchanok2008direct}.

In order to quantify the dynamic effects of a depletion layer,
Donath and coworkers proposed an approximation for the hydrodynamic
friction of a sphere in a non-adsorbing polymer solution by
considering a slip boundary condition at the surface of the particle
\cite{donath1997stokes}. Tuinier and Taniguchi considered a
viscosity profile near the surface that followed polymer segment
density profile and could account for the polymer depletion-induced
flow behavior close to a flat interface \cite{tuinier2005polymer}.
The translational and rotational motion of a sphere and a pair of
spheres through a non-adsorbing polymer were investigated by Fan et.
al.
\cite{tuinier2006depletion,fan2007motion,fan2007asymptotic,fan2010hydrodynamic}
using both a simplified two-layer model
\cite{tuinier2006depletion,fan2007motion} and a continuous viscous
profile \cite{fan2007asymptotic,fan2010hydrodynamic}. For the
continuous case, the equilibrium distribution of polymers was
determined by mean-field theory \cite{fleer2003mean}. When two
colloidal spheres are suspended in a liquid, they transfer momentum
to each other by the hydrodynamic interactions and hence their
stochastic motion is correlated. A popular way to directly measure
the interactions between colloidal particles is to hold the
particles in a desired separation distance which can be done by
applying an optical trap, first introduced by Ashkin et. al.
\cite{ashkin1986observation}. This optical tweezer method has been
used to measure the pair interactions under charge
stabilization~\cite{crocker1994microscopic} and the attractive
entropic effect \cite{verma1998entropic,verma2000attractions}.
Crocker~\cite{crocker1997measurement} developed a blinking optical
trap to measure the free diffusive motion of pair particles. Two
spherical particles were brought into a desired separation distance
when the tweezer is on, while particles start to diffuse freely when
off. The cross-correlation analysis of hydrodynamic interactions of
two particles was studied experimentally using optical
tweezers~\cite{meiners1999direct, bartlett2001measurement,
reichert2004hydrodynamic}. The method has been extended to
investigate the dynamics of optically bounded
particles~\cite{metzger2007measurement} and the shear
effect~\cite{ziehl2009direct,bammert2010dynamics}.

Hence there is a strong need in theoretically quantifying the
polymer depletion effect on the stochastic interactions of a pair of
colloidal particles. Here we present the results of the mobility
functions of two interacting spheres and the thermodynamic potential
due to the entropic effect. The self- and cross-correlation of the
pair particles' random displacements under dilute to semi-dilute
polymer solution conditions are compared to pure solvent case. When
accounting for polymer, the polymer entanglement is neglected and
the fluid flow is assumed Newtonian. Background polymer fluctuations
are also not considered here. The simplified two-layer continuum
model is applied to both hydrodynamic and thermodynamic
interactions. The polymer structure relaxation is assumed much
faster than the long-time Brownian motion such that the depletion
envelop always follows the sphere's trajectory. The mobility is
computed deterministically by the boundary integral method, and the
result is coupled to the random displacements of the colloidal hard
spheres to ensure the consistency of the fluctuation-dissipation
theorem.

\newpage
\section{Theoretical Formulation}
We consider Brownian motion of a pair of isotropic and equal-sized
hard spheres in a dilute to semi-dilute non-adsorbing polymer
solution. It is assumed that each colloidal particle is surrounded
by an assumed equilibrium depletion layer. The mean positions of
both hard spheres are fixed by the optical traps as illustrated in
Fig. 1. The depletion zone around each sphere is represented by the
simplified two-layer model \cite{tuinier2006depletion} that has
uniform solvent viscosity ($\eta_\textrm s$) in the inner layer
($\Omega_\textrm s$) and uniform bulk polymer solution viscosity
($\eta_\textrm p$) in the bulk ($\Omega_\textrm p$).
\\
\begin{figure}[ht]
\includegraphics[trim = 0mm 0mm 0mm 0mm, clip,width=4.5in]{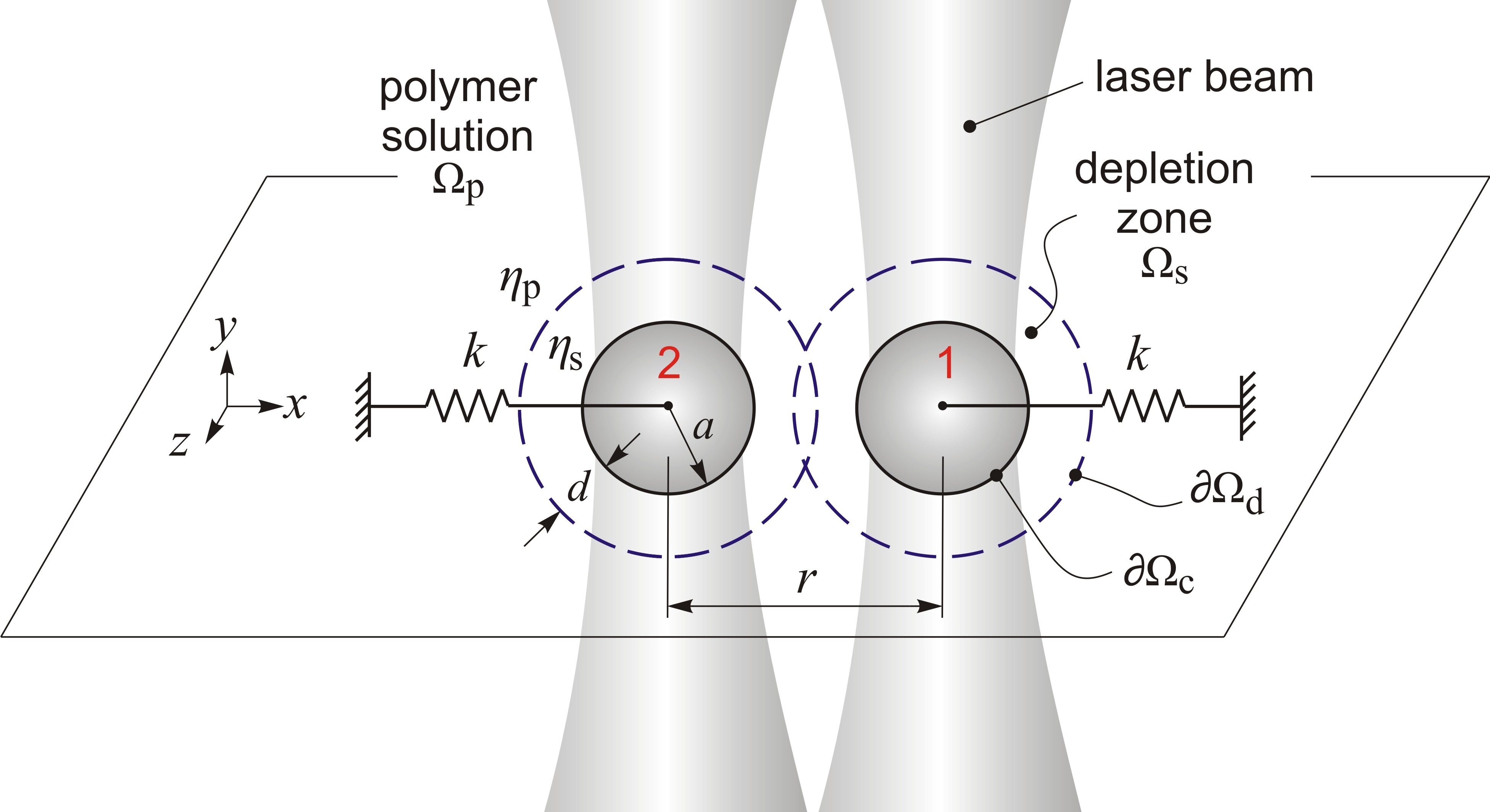}
\caption{Schematic of a pair of interacting colloidal spheres located by optical
tweezers. The domains $\Omega_\textrm p$ and $\Omega_\textrm s$ are for the bulk solution
and the polymer depletion zones, respectively. $\eta_\textrm p$ and $\eta_\textrm s$ are
the corresponding bulk and solvent viscosities. The distance $d$ is the apparent
thickness of the depletion layer, $a$ is colloid radius, and $k$ is the stiffness that
represents the trapping potential. The particle tags 1 and 2 are given.}
\label{F-pair-sch}
\end{figure}

\subsection{Two-Layer Model}
In the bulk, the polymer solution viscosity can be approximated by
the Martin equation~\cite{rodriguez1973graphical}:
\begin{eqnarray}\label{E-Martin}
\lambda=\frac{\eta_\textrm p}{\eta_\textrm
s}=1+[\eta]c_\textrm{b}e^{k_\textrm H [\eta]c_\textrm b}=1+\epsilon
e^{k_\textrm H\epsilon},
\end{eqnarray}
where $\eta_\textrm p$ and $\eta_\textrm s$ are the corresponding
bulk and solvent viscosities, respectively, $c_\textrm b$ is the
bulk polymer concentration, $[\eta]=1/c^*_\textrm b$ is the
intrinsic viscosity of the polymer, $c^*_\textrm b$ is the polymer
overlap concentration, and $k_\textrm H$ is the Huggins coefficient.
For the polymer concentration we use the scaled quantity
$\epsilon=[\eta]c_\textrm b=c_\textrm b/c^*_\textrm b$.

The depletion thickness around the hard spheres can be calculated
based on the bulk concentration and the sizes of the polymers and
colloid spheres. As a correction of the depletion thickness at a
planar surface $d_\textrm P$~\cite{fleer2003mean}, the thickness at
a spherical surface can be expressed as
\begin{eqnarray}\label{E-thk-sphr}
d=a\left[{1+3\frac{d_\textrm P}{a}+\frac{\pi^2}{4}\left(\frac{d_\textrm
P}{a}\right)^2}\right]^{1/3}-a,
\end{eqnarray}
where $d_\textrm P$ is the depletion thickness for a planar surface,
$a$ is the colloid radius, $d_\textrm
P=d_0\sqrt{1/(1+c_1\epsilon^2)}$, where $d_0=2R_g/\sqrt{\pi}$
indicates the thickness in the dilute limit, and $R_g$ is the radius
of gyration of polymers. The coefficient $c_1$ is around $6.02$ for
polymers in a theta solvent
\cite{fleer2003mean,fleer2008analytical}.

According to the Asakara-Oosawa-Vrij (AOV) model
~\cite{asakura1954interaction,vrij1976polymers}, the depletion potential between two hard
spheres in a solution of dilute depletants is given by the product of the osmotic
pressure $\Pi$ and the overlap volume of the polymer depletion
zones~\cite{vrij1976polymers}:
\begin{eqnarray}\label{E-depletioP}
W^{\textrm{dep}}(\epsilon,r)=\begin{cases} \displaystyle\infty~~\textrm{for}~~ r\leq 2a\\
\displaystyle{-\frac{4\pi}{3}(a+d)^3 \left[1-\frac{3}{4}\left(\frac{r}{a+d}\right)
+\frac{1}{16}\left(\frac{r}{a+d}\right)^3\right]
\Pi(\epsilon)}\\~~~~~~~~~~~~\textrm{for}~~ 2a< r\leq 2(a+d)\\
0 \quad\textrm{for}\quad r>2(a+d),
\end{cases}
\end{eqnarray}
where $r=|\mathbf r_1-\mathbf r_2|$ is the center-to-center distance of the hard spheres,
$\Pi(\epsilon)=\Pi_0[1+c_2\epsilon^{2}]$ approximates the osmotic pressure in the dilute
to semi-dilute regime~\cite{fleer2008analytical}, where $\Pi_0(\epsilon)=n_\textrm
b(\epsilon)k_\textrm BT$ is the for dilute limit with $n_\textrm
b(\epsilon)=3\epsilon/(4\pi R_g^3)$ as the number density of polymers, and $c_2$ is
around $4.1$ for a theta solvetnt~\cite{fleer2008analytical}. The corresponding depletion
force therefore is
\begin{eqnarray}\label{E-depletioF}
&&\mathbf F^\textrm{dep}_\alpha(\epsilon,r)= -\nabla
W^\textrm{dep}(\epsilon,r)\\
&&\quad\quad\quad={\pi}(a+d)^2
\left[1-\frac{1}{4}\left(\frac{r}{a+d}\right)^2\right]
\Pi(\epsilon)~\mathbf {\hat r}_{\alpha\beta} \nonumber
\end{eqnarray}
for $2a< r\leq 2(a+d)$, where $\mathbf {\hat
r}_{\alpha\beta}=[\mathbf r_{\beta}-\mathbf r_{\alpha}]/r$ is the
unit vector pointing from particle $\alpha$ to $\beta$.

\subsection{Brownian Interactions}
The translational Brownian motions of both spheres in two quiescent
fluid can be described by the general Langevin
equation~\cite{deutch1971molecular}:
\begin{equation}\label{E-Langevin}
m_\alpha\ddot{\mathbf r}_{\alpha}=\textbf{F}_\alpha^\textrm
H+\textbf{F}_\alpha^\textrm P+\textbf{F}_\alpha^\textrm B
~~~~~~(\alpha=1,2),
\end{equation}
where $m$ is the particle mass, $\alpha$ labels particles 1 and 2,
$\mathbf{r}$ is the particle position vector, and the force
summation acting on the hard spheres includes contributions of
hydrodynamic forces ($\textbf{F}_\alpha^\textrm H$), polymer
depletion and optical trapping forces (non-hydrodynamic,
$\textbf{F}_\alpha^\textrm P$), and the random thermal fluctuation
force ($\textbf{F}_\alpha^\textrm B$) that drives the Brownian
motion. They can be written as
\begin{equation}\label{E-Langevin2a}
\textbf{F}_\alpha^\textrm
H=-\left(\boldsymbol\zeta_{\alpha1}\cdot\dot{\mathbf{ r}}_1
+\boldsymbol\zeta_{\alpha2}\cdot\dot{\mathbf{r}}_2\right),
\end{equation}
\begin{equation}\label{E-Langevin2b}
\textbf{F}_\alpha^\textrm
P=\textbf{F}_\alpha^{\textrm{dep}}+\textbf{F}_\alpha^{\textrm{trap}},
\end{equation}
and
\begin{equation}\label{E-Langevin2c}
\textbf{F}_\alpha^\textrm
B=\boldsymbol\kappa_{\alpha1}\cdot\textbf{x}_1
+\boldsymbol\kappa_{\alpha2}\cdot\textbf{x}_2,
\end{equation}
where $\boldsymbol\zeta$ is the resistance tensor, $\boldsymbol
\kappa$ is the weighting coefficient tensor and the vector
$\textbf{x}$ contains random numbers in Cartesian coordinates and is
defined by a Gaussian distribution with mean $\langle
\textbf{x}_\alpha(t) \rangle=\textbf{0}$ and co-variance $\langle
\textbf{x}_\alpha(t) \textbf{x}_\beta(t+t')
\rangle=2\delta_{\alpha\beta}\delta(t')\textbf{I}$, where $\delta$
is the Kronecker delta, $\textbf{I}$ is the identity tensor, and
$\alpha,\beta$=1, 2. The coefficient $\boldsymbol\kappa$ is related
to the thermal energy $k_\textrm BT$ and the resistance tensor
$\boldsymbol \zeta$ through the fluctuation-dissipation theory
\cite{kubo1966fluctuation},
\begin{equation}
\boldsymbol\zeta_{\alpha\beta} = \frac{1}{k_\textrm BT}
\left(\boldsymbol\kappa_{\alpha1}\cdot
\boldsymbol\kappa_{1\beta}+\boldsymbol\kappa_{\alpha2}
\cdot\boldsymbol\kappa_{2\beta}\right)~~~~~~(\alpha,\beta=1,2).
\end{equation}
The non-hydrodynamic force includes the entropic and optical trapping forces. For a small
displacement away from the equilibrium position of the hard spheres, the optical trap
provides a three-dimensional harmonic force~\cite{Tlusty1998}:
\begin{equation}\label{E-harmonic}
\mathbf F^\textrm{trap}_\alpha=-k\left[\mathbf r_\alpha(t) - \mathbf
r_{\alpha}^{0}\right],
\end{equation}
where $k$ is the apparent stiffness of the potential, and the superscript 0 indicates the
equilibrium position in absence of the depletion potential. The stiffness of the harmonic
force is approximately a linear function of the light intensity \cite{Tlusty1998}.

In order to quantify typical time scales we consider colloidal spheres with radius 500 nm
in an optical trap with a stiffness of 18.5 pN$/\mu$m \cite{meiners1999direct} in an
aqueous solvent. The characteristic relaxation time for the colloidal Brownian motion
under the potential can then be estimated by $6\pi\eta_\textrm s a / k\simeq 10^{-3}$ s.
If the mass density of the colloid is similar to solvent, the momentum relaxation or
decorrelation time, $m/6\pi\eta_\textrm s a$, is about $10^{-7}$ s, where $\eta_\textrm
s$ is the solvent viscosity. The diffusive time scale for the colloid over its radius is
$\sim{6\pi\eta_\textrm{s}a^3}/{k_\textrm BT}\simeq 0.6$ s. Therefore for the long-time
diffusive interaction, the motion of hard spheres is overdamped and the inertial effect
is assumed negligible. By integrating the quasi-steady Langevin equation the new position
for the colloidal sphere $\alpha$ at each simulation time step ($\gg$ momentum
decorrelation time) can be described by the following diffusive displacement
equation~\cite{ermak1978brownian}:
\begin{eqnarray}\label{E-Ermak} \mathbf r_\alpha(t+\Delta
t)=&&\mathbf r_\alpha(t)+\left[\frac{\partial }{\partial \mathbf
r_1(t)}\cdot\mathbf D_{\alpha1}(t)+\frac{\partial }{\partial \mathbf
r_2(t)}\cdot\mathbf D_{\alpha2}(t)\right]\Delta t\\
&&~~+\frac{1}{k_\textrm BT}\left[\mathbf
D_{\alpha1}(t)\cdot\textbf{F}_1^P(t)+\mathbf
D_{\alpha2}(t)\cdot\textbf{F}_2^P(t)\right]\Delta t+\mathbf
R_\alpha(\Delta t),~~\nonumber
\end{eqnarray}
where $\mathbf D_{\alpha\beta}$ ($\alpha,\beta$=1,2) is the diffusion tensor, $\Delta t$
is the integration time step, and $\mathbf R_\alpha$ is the random displacement that has
zero mean $\langle \textbf{R}_\alpha(\Delta t) \rangle=\textbf{0}$, and covariance
$\langle \mathbf R_\alpha(\Delta t)\mathbf R_\beta(\Delta t)\rangle=2\mathbf
D_{\alpha\beta}(t)\Delta t$. The displacements resulting from the spatial variation of
the diffusion tensor, the potential forces acting on both hard spheres, and the thermal
fluctuation are all included in the formulation. Computationally, the random displacement
vector can be determined by $R_i(\Delta t)=\sum_{j=1}^iL_\emph{ij}(\mathbf
D_{\alpha\beta})X_\emph j(\Delta t)$ ($i=1,2,...6$), where $i=1,2,3$ represent Cartesian
components for colloid $\alpha=1$, and $i=4,5,6$ are components for colloid $\alpha=2$,
tensor $L$ is the weighting factor for random displacement, $\mathbf L_{\alpha\beta} =
\left(\mathbf D_{\alpha1}\cdot \boldsymbol\kappa_{1\beta}+\mathbf D_{\alpha2}
\cdot\boldsymbol\kappa_{2\beta}\right)/{k_\textrm BT}$, and $X_1, X_2, ... X_6$ are
series of random variables with zero mean and a covariance of $\sqrt{2\Delta
t}$~\cite{ermak1978brownian}.

\subsection{Mobility Functions}
Since we apply the two layer approach to account for the depletion zones, the
Rotne-Prager tensors~\cite{rotne1969variational} or higher-order
corrections~\cite{batchelor1976brownian} are not applicable for resolving the self- and
mutual-diffusivity. The diffusivity $\mathbf{D}_{\alpha\beta} = k_\textrm
BT\boldsymbol{\zeta}_{\alpha\beta}^{-1}$ here is quantified based on the distance between
the hard spheres ($r$), the thickness of polymer depletion layer ($d$), and the
bulk-to-solvent viscosity ratio ($\lambda=\eta_\textrm{p}/\eta_\textrm{s}$). The
self-diffusivity can be formulated as a correction of the mobility
functions~\cite{batchelor1976brownian,dhont1996introduction} for the motion of a pair of
spheres in a homogeneous medium,
\begin{eqnarray}\label{E-diff-ten-1}
&&\mathbf D_{\alpha\alpha}(r,d,\lambda)= \frac{D_0}{g_0(d,\lambda)}\Big[\mathbf
I+A^\textrm {s}(r,d,\lambda)\mathbf{\hat r}_{\alpha\beta}\mathbf{\hat
r}_{\alpha\beta}+B^\textrm {s}(r,d,\lambda)\left(\mathbf I - \mathbf{\hat
r}_{\alpha\beta}\mathbf{\hat r}_{\alpha\beta}\right) \Big],
\end{eqnarray}
where $\alpha,\beta=1,2$ and $\alpha\neq\beta$, $g_0(d,\lambda)$ is the correction factor
for the hydrodynamic friction coefficient of an isolated hard sphere $6\pi\eta_\textrm s
a$ due to the depletion layer \cite{fan2007motion}, $D_0=k_\textrm BT/6\pi\eta_\textrm
sa$ is the diffusivity of an isolated sphere in a pure solvent, $\mathbf I$ is the
identity matrix, $A$ and $B$ are mobility functions in parallel and perpendicular
directions, respectively, and the superscript s is for self-mobility. Throughout this
paper the correction factor $g$ with any super- and sub-script describes the deviation of
the friction from a single sphere in a pure solvent (for which $g=1$). Similarly, the
mutual diffusivity can be expressed as
\begin{eqnarray}\label{E-diff-ten-2}
&&\mathbf D_{\alpha\beta}(r,d,\lambda)=\frac{D_0}{g_0(d,\lambda)}\Big[A^\textrm
{c}(r,d,\lambda)\mathbf{\hat{r}}_{\alpha\beta}\mathbf{\hat{r}}_{\alpha\beta}+B^\textrm
{c}(r,d,\lambda)\left(\mathbf I -
    \mathbf{\hat{r}}_{\alpha\beta}\mathbf{\hat{r}}_{\alpha\beta}\right)\Big],
\end{eqnarray}
where $\alpha,\beta=1,2$ and $\alpha\neq\beta$, the superscript c indicates
cross-mobility. The analytical expression for a single particle based on the two-layer
approximation is given by~\cite{fan2007motion}
\begin{eqnarray}\label{E-correction}
&&g_0(d,\lambda)=\frac{2}{\Gamma}
\left[\left(2+{3}{\lambda}^{-1}\right)
\left(1+{d}^*\right)^6-2\left(1-{\lambda}^{-1}\right)
\left(1+{d}^*\right)\right],
\end{eqnarray}
where $d^*=d/a$ is the normalized depletion thickness and
\begin{eqnarray}
\Gamma=&& 2\left(2+{3}{\lambda}^{-1}\right) \left(1+{d}^*\right)^6 -
3\left(3+{2}{\lambda}^{-1}\right)
\left(1-{\lambda}^{-1}\right)\left(1+{d}^*\right)^5\\&&+10\left(1-{\lambda}^{-1}\right)\left(1+
{d}^*\right)^3-9\left(1-{\lambda}^{-1}\right)
\left(1+{d}^*\right)+4\left(1-{\lambda}^{-1}\right)^2.\nonumber
\end{eqnarray}
Here the mobility functions are decoupled and computed by the boundary integral method.
Specifically, the four functions are expressed as
\begin{eqnarray}\label{E-mobilityFa-1}
&&\displaystyle{A^\textrm {s}(r,d,\lambda)=
\frac{g_0(d,\lambda)}{2}\left[\frac{1}{g_{\|}^{\textrm
I}(r,d,\lambda)} + \frac{1}{g_{\|}^{\textrm
{II}}(r,d,\lambda)}\right]-1},~~~~~~\\
&&\displaystyle{B^\textrm {s}(r,d,\lambda) =
\frac{g_0(d,\lambda)}{2} \left[\frac{1}{g_\bot^\textrm
I(r,d,\lambda)} +
\frac{1}{g_\bot^\textrm{II}(r,d,\lambda)}\right]-1},\nonumber\\
&&\displaystyle{A^\textrm {c}(r,d,\lambda) =
\frac{g_0(d,\lambda)}{2}\left[\frac{-1}{g_\|^\textrm{I}(r,d,\lambda)}
+ \frac{1}{g_\|^\textrm{II}(r,d,\lambda)}\right]},\quad\textrm{
 and}\nonumber\\
&&\displaystyle{B^\textrm {c}(r,d,\lambda) =
\frac{g_0(d,\lambda)}{2}\left[\frac{-1}{g_\bot^\textrm
I(r,d,\lambda)} +
\frac{1}{g_\bot^\textrm{II}(r,d,\lambda)}\right]},\nonumber
\end{eqnarray}
where $g_{\|}^\textrm I$, $g_{\|}^\textrm {II}$, $g_{\bot}^\textrm I$, and
$g_{\bot}^\textrm {II}$ are the corrections or the scaled resistances due to depletion
effect and the hydrodynamic interaction between both spheres, defined as
\begin{eqnarray}\label{E-corrections-1}
&g_{\|}^\textrm I(r,d,\lambda)=\displaystyle\frac{F_{\|}^\textrm
I(r,d,\lambda)}{6\pi\eta_\textrm s aU}, ~~ g_{\|}^\textrm
{II}=\displaystyle\frac{F_{\|}^\textrm {II}}{6\pi\eta_\textrm s aU}, &\\&g_{\bot}^\textrm
I=\displaystyle\frac{F_{\bot}^\textrm I}{6\pi\eta_\textrm s aU},
~~\textrm{and}~~~~g_{\bot}^\textrm {II}=\displaystyle\frac{F_{\bot}^\textrm
{II}}{6\pi\eta_\textrm s aU},\nonumber&
\end{eqnarray}
where the four hydrodynamic interaction modes $F_{\|}^\textrm I$,
$F_{\|}^\textrm {II}$, $F_{\bot}^\textrm I$, and $F_{\bot}^\textrm
{II}$ (see Fig. \ref{F-Hyd-int} in the Results and Discussion) are
the computed resistance on each sphere for the interaction parallel
and perpendicular to the center-to-center line, respectively. $U$ is
the velocity magnitude of both colloids, and $g\rightarrow1$ as
$r\rightarrow\infty$. For mode I (indicated by the superscript) both
spheres move in the same direction, whereas in mode II both spheres
move in the opposite direction. Accordingly, the divergence of the
diffusivity tensor becomes (Appendix A)
\begin{eqnarray}\label{E-graddiff-1}
&\displaystyle{\frac{\partial~~ }{\partial \mathbf r_1}\cdot \mathbf
D_{11}(r,d,\lambda) = \frac{-D_0}{g_0}\left[ \frac{\partial
A^\textrm {s}}{\partial r}+\frac{2(A^\textrm s-B^\textrm
s)}{r}\right]\mathbf{\hat{r}}_{12}= -\frac{\partial~~}{\partial
\mathbf r_2}\cdot\mathbf D_{22},}&
\end{eqnarray}
and
\begin{eqnarray}\label{E-graddiff-2}
\displaystyle{\frac{\partial~~}{\partial \mathbf r_2}\cdot\mathbf
D_{12}(r,d,\lambda) =\frac{D_0}{g_0}\left[\frac{\partial A^\textrm
{c}}{\partial r}+\frac{2(A^\textrm c-B^\textrm
c)}{r}\right]\mathbf{\hat{r}}_{12}= -\frac{\partial~~}{\partial
\mathbf r_1}\cdot\mathbf D_{21}.}
\end{eqnarray}

For a pair of hard spheres moving in a pure solvent the analytical
approximation for all modes of hydrodynamic interactions have been
provided by Stimson and Jeffery ~\cite{Stimson1926},
Brenner~\cite{brenner1961slow}, and
O'Neill~\cite{o1970asymmetrical}. However, the analytical results
that account for the polymer depletion effect on pair interaction
are not available. Here we apply the boundary integral method to
compute the hydrodynamic interactions, i.e., $F_{\|}^\textrm I$,
$F_{\|}^\textrm {II}$, $F_{\bot}^\textrm I$, and $F_{\bot}^\textrm
{II}$ in order to determine the random displacements that are
consistent with the fluctuation-dissipation theorem.

\subsection{Integral Formulation of the Pair Hydrodynamic Interaction}

The quasi-steady Stokes flow applied to the two-layer model can be formulated as
\begin{eqnarray}\label{E-Stokes-1}
\eta_\text p\nabla^2\boldsymbol{\mathit{v}}^{(\textrm p)}-\nabla p^{(\textrm p)}=0,~
\nabla \cdot \boldsymbol{\mathit v}^{(\textrm p)} = 0 ~~~\textrm{for}~~~\mathbf
r\in\Omega_\textrm p,
\end{eqnarray}
and
\begin{eqnarray}\label{E-Stokes-2}
\eta_\text s\nabla^2\boldsymbol{\mathit v}^{(\textrm s)}-\nabla p^{(\textrm s)}=0,~
\nabla \cdot \boldsymbol{\mathit v}^{(\textrm s)}=0~~~\textrm{for}~~~\mathbf
r\in\Omega_\textrm s,
\end{eqnarray}
where $\boldsymbol{\mathit v}$ is velocity, $p$ is pressure, $\mathbf r$ is position
vector, superscript $\textrm{s}$ and $\textrm{p}$ indicate the solvent (depletion zone)
and bulk polymer solution, respectively. The no-slip boundary condition is applied at the
particle surface by combining the translational and rotational velocity,
$\boldsymbol{\mathit v}^\textrm{(s)}=\mathbf U_\alpha \boldsymbol +
\boldsymbol\omega_\alpha\times\boldsymbol\ell_\alpha$ ($\alpha=1,2$). The far-field
boundary conditions are $\boldsymbol{\mathit v}^{\text {(p)}}\to \mathbf 0$ and
$p^\textrm{(p)}\to p^\infty$ as $\mathbf r\to \infty$. At the interface between the
depletion zone and the bulk polymer solution, the velocity and stress are continuous,
i.e., $\boldsymbol{\mathit v}^{\text {(s)}} = \boldsymbol{\mathit v}^{\text {(p)}}$, and
$\boldsymbol\tau^{\text {(s)}} = \boldsymbol \tau^{\text {(p)}}$ presuming that the
surface tension at the interface is negligible. The translation velocity for four
interactive modes is defined as
\begin{eqnarray}\label{E-BC}
&& \|\textrm{ mode I}:~~\mathbf U_1=\mathbf U_2=U_0~\mathbf{\hat e}_x, \\
&& \|\textrm{ mode II}:~~\mathbf U_1=-\mathbf U_2=-U_0~\mathbf{\hat e}_x,  \nonumber \\
&& \bot\textrm{ mode I}:~~\mathbf U_1=\mathbf U_2=U_0~\mathbf{\hat e}_y,~\textrm{and } \nonumber \\
&& \bot\textrm{ mode II}:~~\mathbf U_1=-\mathbf U_2=-U_0~\mathbf{\hat e}_y \nonumber,
\end{eqnarray}
where $U_0$ is the velocity magnitude and both spheres are aligned along the $x$-axis.
These modes are illustrated in Fig. \ref{F-Hyd-int}.

\begin{figure}[ht]
\begin{center}
\includegraphics[trim = 0mm 0mm 0mm 0mm, clip, width=2.75in]{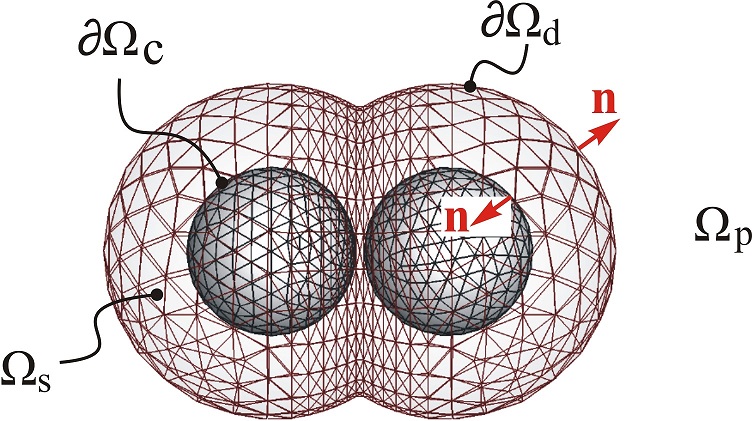}
\end{center}
\caption{Illustration of the surface mesh applied to the integral
computation. $\partial\Omega_\textrm d$ is the interface between the
depletion zone and the bulk polymer solution. The mesh is refined
near the region of overlapping depletion zones.}\label{F-mesh}
\end{figure}

To compute the resistance, the integral formulation of the Stokes
flow \cite{ladyzhenskaya1969mathematical} is applied for the
two-layer model. In the fluid domain $\Omega$ bounded by surface
$\partial\Omega$, the velocity field satisfies the integral momentum
equation, $\int_{\partial\Omega}f_\emph i
G_\emph{ij}dS-\eta\int_{\partial\Omega} v_\emph i
T_\emph{ijk}n_\emph kdS=0$ when the source point is located outside
the fluid domain, while $v_\emph j= -\int_{\partial\Omega}f_\emph i
G_\emph{ij}/(8\pi\eta)dS+\int_{\partial\Omega} v_\emph i
T_\emph{ijk}n_\emph k/(8\pi)dS$ when the source point is within the
domain. Here $\eta$ is the viscosity, $i$ and $j$ are Einstein
notations, $f_\emph i=\tau_{ij}n_j$ is traction, $n$ represents the
surface normal pointing into the fluid, $G_\emph{ij}$ is the
fundamental solution (Stokeslet) of the Stokes equation, and
$T_\emph{ijk}$ is its corresponding stress field (stresslet),
written as
\begin{equation}\label{stokeslet stresslet}
    G_\emph{ij}(\mathbf{r},\mathbf{r}_0)=\frac{
    \delta_\emph{ij}}{x}+\frac{x_\emph i x_\emph
    j}{x^3},~~~\textrm{and}~~~
    T_\emph{ijk}(\mathbf{r},\mathbf{r}_0)=-\frac{6 x_\emph i x_\emph j x_\emph
    k}{x^5},
\end{equation}
here $\mathbf r_0$ and $\mathbf r$ are the source and field points,
respectively, $x=|\mathbf x|$ and $ \textbf{x}=\mathbf r-\mathbf
r_0$. By applying the integral formulation and incorporating the
stress-free condition at the interface between the depletion zone
and the bulk solution $\partial\Omega_\textrm d$ (Fig.
\ref{F-mesh}), $\Delta f_\emph i=f^{(\textrm p)}_\emph i-f^{(\textrm
s)}_\emph i=0$, we have
\begin{eqnarray}\label{E-BI-dim1}
&&v_\emph j(\mathbf r_0\in \Omega_p) = \frac{1}{8\pi\eta_\text
p}\small\int_{\partial\Omega_\textrm c}f_\emph i G_\emph{ij}dS
\\
&&~~~~-\frac{1}{8\pi\lambda}\int_{\partial\Omega_{\textrm c}}
({U_\emph i}+\varepsilon_\emph{ijk}{\omega_\emph j} {\ell_\emph k})
T_\emph{ijk}n_\emph k
dS+\frac{1-1/\lambda}{8\pi}\int_{\partial\Omega_\textrm d} v_\emph i
T_\emph{ijk}n_\emph kdS\nonumber
\end{eqnarray}
and
\begin{eqnarray}\label{E-BI-dim2}
&&\frac{1}{\lambda} v_\emph j(\mathbf r_0\in \Omega_s) =
\frac{1}{8\pi\eta_\text p}\int_{\partial\Omega_\textrm c}f_\emph i
G_\emph{ij}dS\\
&&~~-\frac{1}{8\pi\lambda}\int_{\partial\Omega_{\textrm c}}
({U_\emph i}+\varepsilon_\emph{ijk}{\omega_\emph j} {\ell_\emph k})
T_\emph{ijk}n_\emph k
dS+\frac{1-1/\lambda}{8\pi}\int_{\partial\Omega_\textrm d} v_\emph i
T_\emph{ijk}n_\emph kdS,\nonumber
\end{eqnarray}
where $\varepsilon_\emph{ijk}$ is the permutation tensor, and the
directions of surface normals are given in Fig. \ref{F-mesh}. As the
source approaches the colloidal surface and the depletion interface,
the Cauchy principle value is applied to the double-layer integral
that includes the velocity term \cite{pozrikidis1992boundary}.

\noindent As a result, the corresponding integral equation can be
derived for the two-layer model for the entire computation domain
shown in Fig. \ref{F-mesh}, expressed as
\begin{eqnarray}\label{E-BI-dim}
&&\mathcal H v_\emph j(\mathbf r_0)=\frac{1}{8\pi\eta_\text
p}\int_{\partial\Omega_\textrm c}f_\emph i G_\emph{ij}dS
\\
&&~~-\frac{1}{8\pi\lambda}\int_{\partial\Omega_{\textrm c}}
({U_\emph i}+\varepsilon_\emph{ijk}{\omega_\emph j} {\ell_\emph k})
T_\emph{ijk}n_\emph k
dS+\frac{1-1/\lambda}{8\pi}\int_{\partial\Omega_\text d} v_\emph i
T_\emph{ijk}n_\emph kdS,\nonumber
\end{eqnarray}
where coefficient $\mathcal H=1$ for $\mathbf r_0\in\Omega_\textrm p$, $\mathcal
H=1/\lambda$ for $\mathbf r_0\in\Omega_\textrm s$, $\mathcal H =
({\lambda+1})/({2\lambda})$ for $\mathbf r_0\in\partial\Omega_\textrm d$, and $\mathcal
H={1}/({2\lambda})$ for $\mathbf r_0\in\partial\Omega_\textrm c$. In the integral
equation the rotational velocity $\omega_\emph j$ is unknown \textit{a priori}. It can be
determined by the vanishing torque applied to both isotropic spheres due to the
hydrodynamic coupling,
\begin{equation}\label{E-rotat}
\mathcal T_\alpha=\left|\int_{\partial\Omega_{\textrm
c,\alpha}}\varepsilon_\emph{ijk}{\ell_\emph j}{f_\emph k} dS\right|=0.
\end{equation}
In summary, the integral equations (Eqs. \ref{E-BI-dim} and
\ref{E-rotat}) are discretized and computed for the primary
variables $v_j(\mathbf r_0\in\partial\Omega_\textrm d), f_j(\mathbf
r_0\in\partial\Omega_\textrm c)$, and the rotational velocity for
each sphere $\omega_j$. Once the boundary values are found, the
results in the fluid domains can be obtained from Eqs.
(\ref{E-BI-dim1}) and (\ref{E-BI-dim2}).

\newpage

\section{Results and discussion}

\begin{figure}[ht]
\includegraphics[width=6.4in]{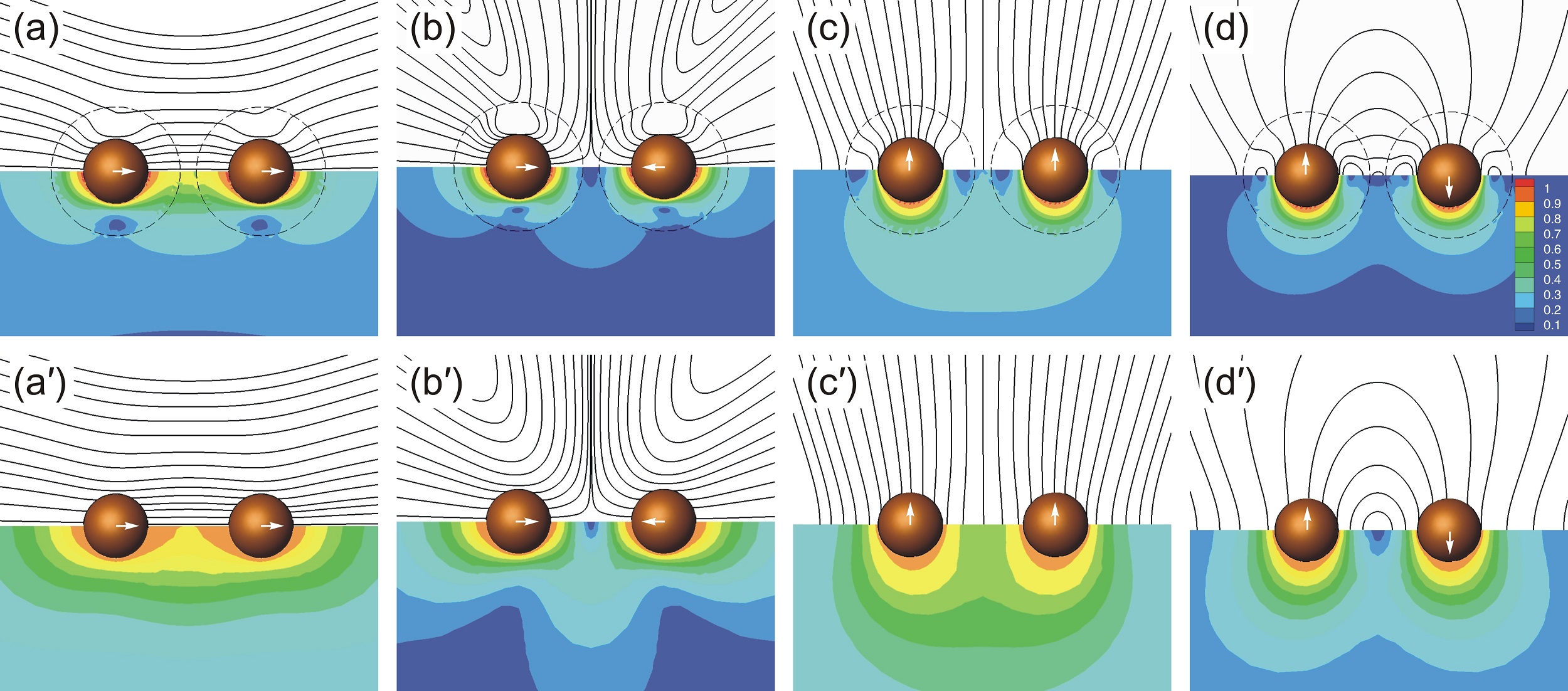}
\caption{Streamlines and contours of the normalized velocities
around interacting spheres in parallel and transverse directions,
with (top panels from a to d) and without  (a$'$ to d$'$) polymer
depletion zones. Parameters used for the demonstration are: $d/a=1$,
$r/a=4.5$, and $\lambda=10$.} \label{F-Hyd-Post}
\end{figure}

\begin{figure}[ht]
\includegraphics[width=6.2in]{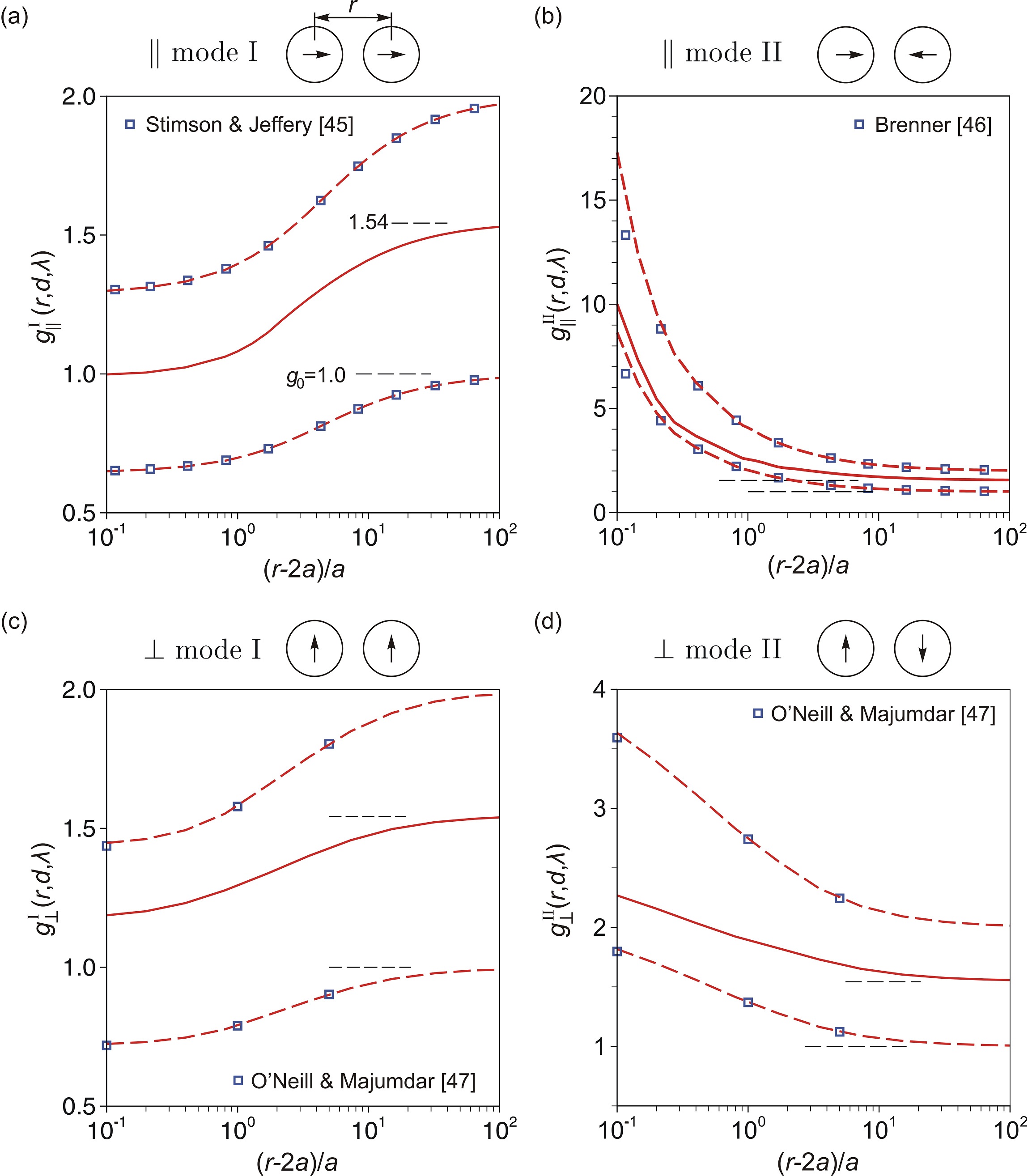}
\caption{Semi-log plots showing the hydrodynamic correction function
$g$ versus the scaled separation distance for both parallel and
perpendicular motions to the center-to-center line along the same
(a, c) and opposite (b, d) directions. The square data points are
the exact series solution for a uniform solvent (lower bound) and
uniform bulk solution (upper
bound)~\cite{Stimson1926,brenner1961slow,o1970asymmetrical}. The
dashed-lines indicate boundary integral results as a numerical
validation. The solid curves show the results for hard spheres in a
nonadsorbing polymer solution with depletion layer described using
the two-layer model.} \label{F-Hyd-int}
\end{figure}

The stochastic motion of both spheres are correlated due to
hydrodynamic and thermodynamic interactions. The four hydrodynamic
modes represented by $F_{\|}^\textrm I$, $F_{\|}^\textrm {II}$,
$F_{\bot}^\textrm I$, and $F_{\bot}^\textrm {II}$ are in general
functions of the separation distance $r$, the polymer-to-sphere size
ratio, and the polymer concentration. The latter quantity both
affects the bulk viscosity as well as the strength of the depletion
attraction. Here the two-layer model requires two input parameters,
the depletion thickness $d$ and the bulk-to-solvent viscosity ratio
$\lambda$. Figure~\ref{F-Hyd-Post} shows the flow patterns induced
by the moving spheres corresponding to two parallel
(\ref{F-Hyd-Post}a and \ref{F-Hyd-Post}b) and two perpendicular
(\ref{F-Hyd-Post}c and \ref{F-Hyd-Post}d) modes with respect to the
center-to-center line. Patterns 3a$'$ to 3d$'$ are corresponding
typical Stokes flow for comparison. In the very dilute limit, all
flow patterns are similar to the cases in a uniform fluid medium as
expected (not shown here), while for a higher value of viscosity
ratio, e.g. $\lambda=10$ as shown in Fig. \ref{F-Hyd-Post}b and
\ref{F-Hyd-Post}d, circulations appear in the depletion zone near
the particle surface due to the cage-like behavior, which is similar
to what we reported earlier for the single particle
\cite{tuinier2006depletion,fan2007motion}. The near-field effect has
significant impact on the stress distribution and therefore the
overall resistance applied to the spheres. The circulation further
complicates the slip-like behavior and changes the shear and normal
viscous force near the front and aft surface. The overall slip-like
behavior yields a fast decay of the velocity magnitude away from the
particle surface. Overall, mode II has a great reduction of the
resistance, whereas in mode I the reduction is less significant. As
later shown in Figs. \ref{F-Hyd-int}, the hydrodynamic correction
factor approaches the resistance-distance curve dominated by the
solvent viscosity (the lower bound) in modes II for both parallel
and perpendicular directions.

Figures \ref{F-Hyd-int}(a) and \ref{F-Hyd-int}(b) show the
correction factors (Eq. \ref{E-corrections-1}) for the hydrodynamic
resistance at various scaled separation distance under parallel
motion along the same (a) and opposite (b) directions. The boundary
integral result is validated by the analytical approximations (the
upper and lower bounds) for mode I~\cite{Stimson1926} and mode
II~\cite{brenner1961slow} written as
\begin{eqnarray}\label{E-Stimson}
g_\|^\textrm
I=&&\frac{4}{3}\sinh(\theta)\sum_{n=1}^\infty\frac{n(n+1)}{(2n-1)(2n+3)}\\
&&\times\left\{1-\frac{4\sinh^2(n+\frac{1}{2})\theta-(2n+1)^2\sinh^2\theta}
{2\sinh(2n+1)\theta+(2n+1)\sinh2\theta} \right\},
\end{eqnarray}
where $\theta=\cosh^{-1}({r}/{2a})$ and
\begin{eqnarray}\label{E-Brenner}
g_\|^\textrm
{II}=&&\frac{4}{3}\sinh(\theta)\sum_{n=1}^\infty\frac{n(n+1)}{(2n-1)(2n+3)}\\
&&\times\left\{\frac{4\cosh^2(n+\frac{1}{2})\theta+(2n+1)^2\sinh^2\theta}
{2\sinh(2n+1)\theta-(2n+1)\sinh2\theta} -1\right\}.
\end{eqnarray}
In the presence of the depletants, the hydrodynamic resistance is
bounded by two limiting cases: solvent only and uniform polymer
solution without the depletion effect (shown by the analytical data
points and numerical dashed curves). The solid curve in between the
upper and lower bounds illustrates the numerical results obtained
for hard spheres in a non-adsorbing polymer solution described using
the two-layer model. The asymptote $g\rightarrow1.54$ is the
analytical result for the resistance of a single sphere under the
same depletion condition~\cite{fan2007motion}, and $g\rightarrow1.0$
is the Stokes limit. For the uniform polymer solution the asymptote
is $g\to\lambda$=2 (the upper bound). In the lubrication regime when
two spheres are close to each other ($r/a\to2$, mode II, 4b and 4d),
the resistance approaches the solvent limit due to the dominant
stress contribution from the polymer-depleted liquid film in between
the spheres. However, on mode I (4a and 4c) the thin liquid film has
much less influence on the resistance compared with that from the
surrounding fluid. The analytical results on the perpendicular modes
(square data points in Figs. 4c and 4d) are provided by O'Neill and
Majumdar~\cite{o1970asymmetrical}, which are presented by the lower
and upper bounds that we also used to validate the boundary integral
results. The three data points in Fig. \ref{F-Hyd-int}c and d The
same asymptotes as presented in the parallel mode are given for
large separation distance. Note that in the lubrication limit, the
liquid film has less shear effect on mode II compared with the
parallel motion due to the torque-free condition.

\newpage

\begin{figure}[ht]
\begin{center}
\includegraphics[width=6.2in]{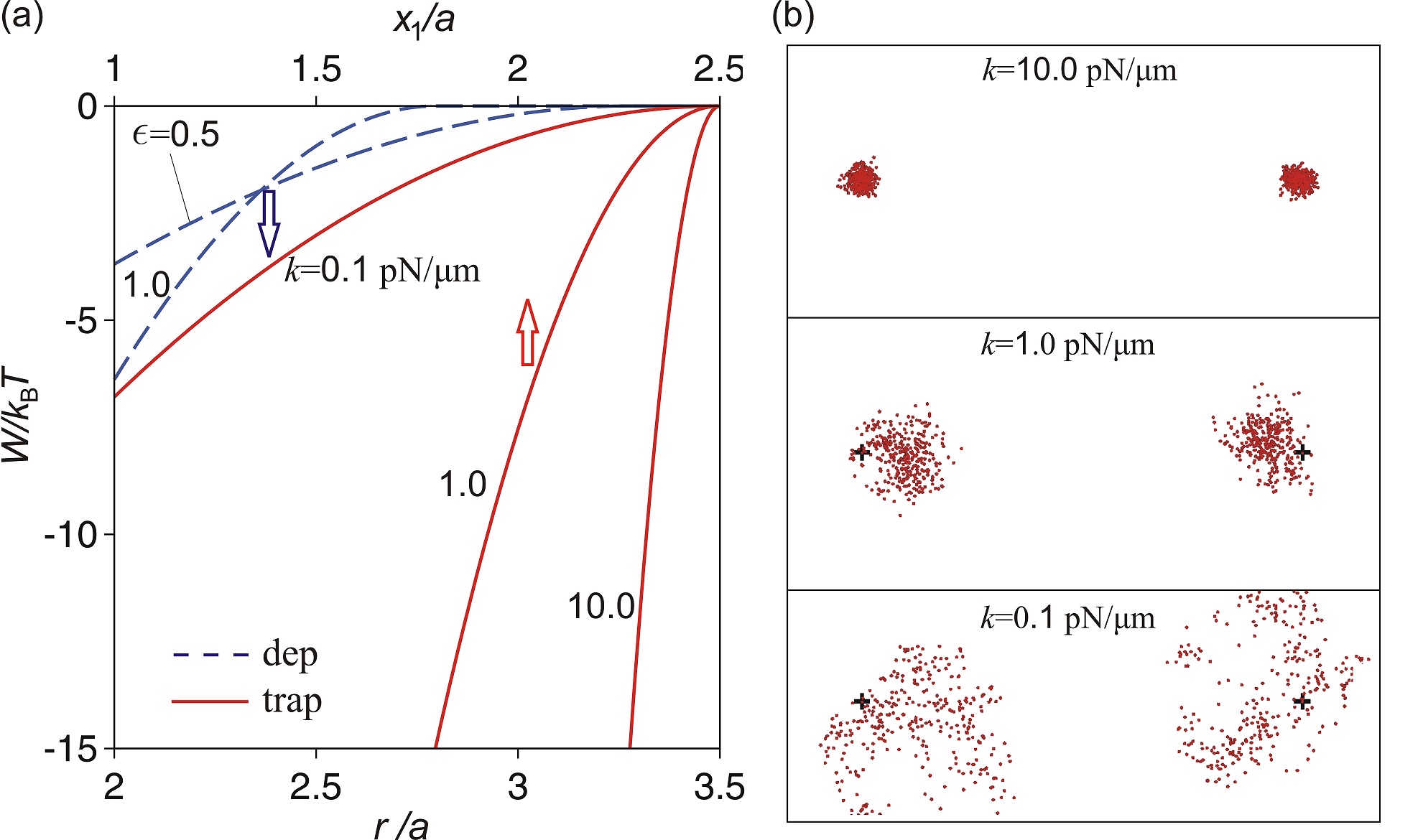}
\end{center}
\caption{(a) The comparison of the potential of the optical trap
applied to the particle with the trapping center located at
$x_1/a=2.5$ and the polymer depletion potential between two spheres
versus the center-to-center separation distance $r/a$ (=2 when
spheres are in contact with each other). (b) A test run of Brownian
simulations with data points showing the particle position per 200
time steps. The total simulation time is 2 s. In this test case,
$d/a=1$ and $\lambda=2$. The centers of two traps are indicated by
the plus signs.} \label{F-potential}
\end{figure}
In Fig. \ref{F-potential} we compare typical depletion and trapping
harmonic potentials applied to the hard spheres. In general the
attractive force is enhanced by a higher polymer concentration if
the overlap volume remains the same. However, the depletion
thickness reduces as the bulk polymer concentration increases, which
reduces the overlap volume and suppresses the attractive force. The
value for the stiffness of the optical trap has a significant caging
effect on the Brownian motion of both spheres as the simulation
results illustrated in Fig. \ref{F-potential}(b). Parameters used in
the simulation are: time step $\triangle t=$28.5 $\mu$s, radius
$a=500$ nm, polymer concentration $\epsilon=$ 0.5, and depletion
thickness $d/a=0.69$ that corresponds to $R_g/a=1$. The distance
between the traps and the initial particle separation distance are
set to $r/a$=2.5. Both hydrodynamic and depletion effects are taken
into account, the data points clearly shows the attractive depletion
effect on the probability distribution of the spheres, with the
characteristic distribution length inversely proportional to the
square root of $k$. Next we consider the correlation analysis that
better characterizes the ensemble results.

As a baseline comparison for the correlation analysis, the long-time
autocorrelation of the random displacement is considered for a
single Brownian sphere with and without polymer depletion, which can
be formulated by the analytical result~\cite{doi1988theory}:
\begin{eqnarray}\label{E-corr-iso}
\langle x(t)x(0)\rangle = \frac{k_\textrm
BT}{k}\left[\exp\left(-\frac{t}{\tau} \right)\right],
\end{eqnarray}
\noindent where ${k_\textrm BT}/{k}$ measures the thermal energy
versus the strength of the harmonic potential, and more importantly
the decorrelation time depends on corrected hydrodynamic resistance,
and the harmonic potential,
\begin{equation}\label{E-dec-tim}
\tau=\frac{6\pi\eta_\textrm s ag_0(\lambda,d)}{k},
\end{equation}
and the mean square displacement can also be formulated as
\begin{eqnarray}\label{E-msd-iso}
\langle x(t)^2\rangle = \frac{k_\textrm
BT}{k}\left[1-\exp\left(-\frac{2t}{\tau} \right)\right].
\end{eqnarray}
For spheres with a radius of 500 nm and an optical trap with
stiffness $k=$10, 1, and 0.1 pN/$\mu$m, the corresponding
characteristic length squares for the autocorrelation function
$\sqrt{k_\textrm BT/k}$ are approximately 20, 65, and 200 nm in a
pure solvent. The corresponding decorrelation times are 1, 10, and
100 ms, respectively, whereas in a dilute polymer solution, the
decorrelation times increases slightly to 1.33, 13.3, and 133 ms in
a pure solvent.

This increase is due to the enhanced viscous resistance or slower
motion of the particles. Although the discussion here is applicable
to an isolated particle, the resulting characteristic distribution
length is about the same order of magnitude as the diffusive length
(independent of hydrodynamic interactions) observed in Fig. 5b
because the total simulation time (2 s) is far beyond the
decorrelation time. In other words, the depletion effect shifts the
probability distribution of the random walk instead of altering the
characteristic diffusive length. If the bulk viscosity is used
instead of the two-layer depletion model, the decorrelation times
would be 1.53, 15.3, and 153 ms, respectively. One can speculate
that the result is applicable for colloid-polymer dispersions with
colloid volume fraction $\ll10^{-5}$ or an averaged pair separation
distance $\gg 100a$.

\begin{figure}[ht]
\begin{center}
\includegraphics[width=6.2in]{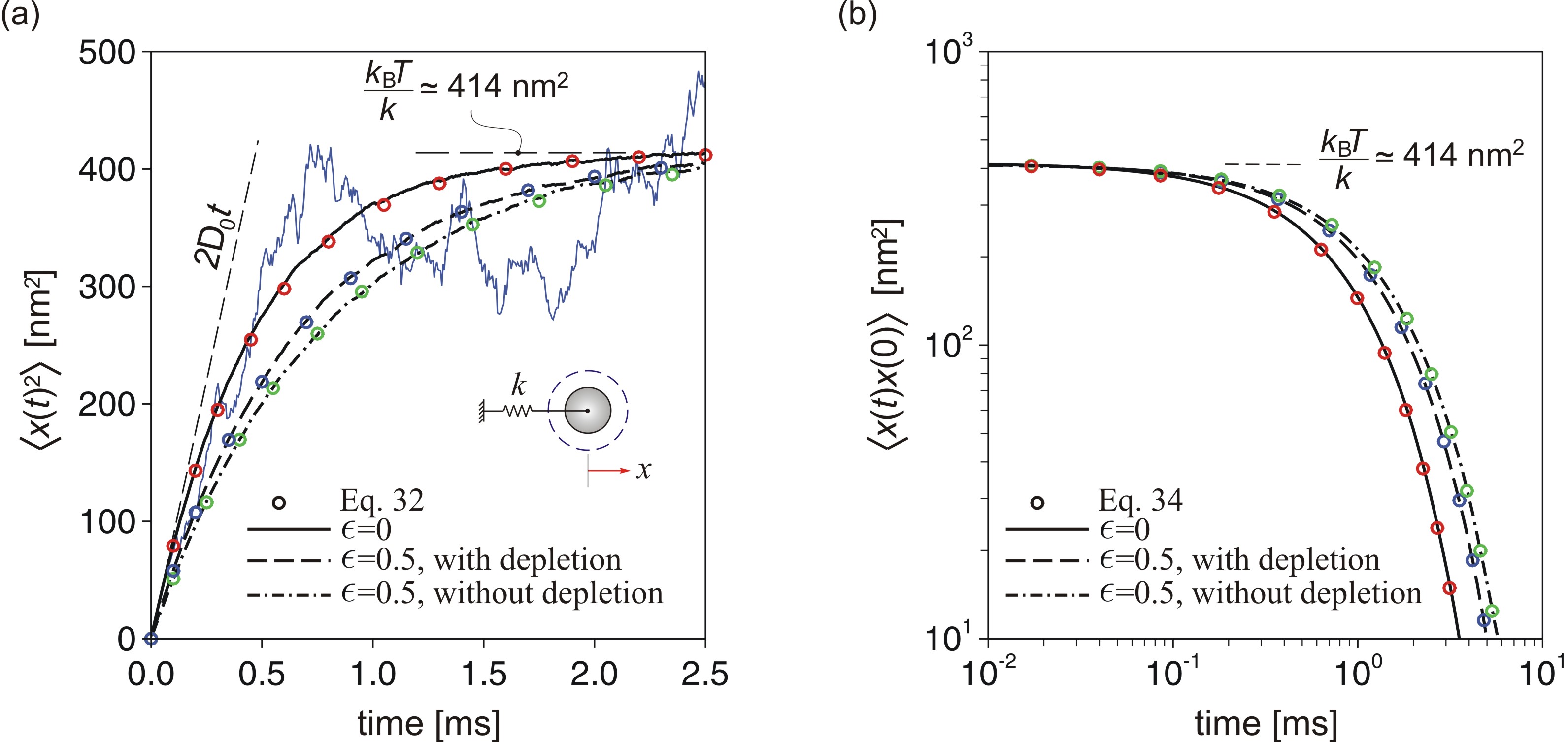}
\end{center}
\caption{Brownian simulation results for a single sphere showing the
mean square displacement versus time (a) and the long-time
autocorrelation function (b) of the random motion of the isolated
sphere in pure solvent ($\epsilon=0$) and dilute polymer solution
($\epsilon=0.5$) with and without the polymer depletion effect under
ambient temperature. The thin blue curve is the average from 100
samples, and the thick black curve is the result from averaging over
10$^4$ samples .} \label{F-single-auto}
\end{figure}
In Fig. \ref{F-single-auto} results are shown for the random motion
of a tapped single sphere. In Fig. 6a the mean square displacement
is plotted versus time in a pure solvent and in polymer solutions
with and without depletion. In Fig. \ref{F-single-auto}a, the thin
blue curve is the ensemble average over 100 samples which is
considerably noisy compared to the thick black curve from the
average over 10$^4$ samples. Parameters applied to the simulations
are: $a$=500 nm, $k$=10 pN/$\mu$m, $R_g/a$=1, viscosity ratio
$\lambda$=1.611, depletion thickness $d/a$=0.638 for the normalized
polymer concentration $\epsilon=0.5$, and the time step for the
Brownian simulation $\Delta t=$3$\times10^{-5}$ s. In Fig.
\ref{F-single-auto}b, the data points represent the analytical
solution and the lines are from Brownian simulations. The sample
size for computing the autocorrelation function is sufficiently
large by considering total simulation time around 10 s. The decay
time can be obtained from the best fit to the numerical results
based on the exponential decay given by Eq. \ref{E-dec-tim}. The
computed versus analytical decay times are 0.9424, 1.3298, and
1.5183 ms for $\epsilon=0$, $\epsilon=0.5$ with depletion, and
$\epsilon=0.5$ without depletion, respectively. Under the same
stiffness, the large viscous effect essentially increases the
decorrelation time. At shorter times the difference is negligible,
whereas at longer times the viscous effect enhances the
autocorrelation along the harmonic force direction as expected.
Therefore, the case without the depletion effect has the largest
decorrelation time (and thus the $g$ factor) due to the highest
viscous resistance.

For the pair interaction, the analytical pair correlation function of spheres in
homogeneous fluids is available~\cite{bartlett2001measurement,meiners1999direct}. A
straightforward extension of this result to incorporate the depletion effect for the
two-dimensional movement of a pair of spheres in $x-$ and $z-$direction ($y$ and $z$ are
equivalent) can be written as
\begin{eqnarray}\label{E-pairmob}
&&\begin{bmatrix}
\dot{x}_1 \\
\dot{z}_1 \\
\dot{x}_2 \\
\dot{z}_2
\end{bmatrix}=
\begin{bmatrix}
1+A^\textrm{s} & 0 & A^\textrm{c} & 0\\
0 & 1+B^\textrm{s} & 0 & B^\textrm{c}\\
A^\textrm{c} & 0 & 1+A^\textrm{s} & 0\\
0 & B^\textrm{c} & 0 & 1+B^\textrm{s}
\end{bmatrix}
\begin{bmatrix}
-kx_1-F^\textrm{dep}+F^{\textrm B}_1\\
-kz_1+F^{\textrm B}_1\\
-kx_2+F^\textrm{dep}+F^{\textrm B}_2\\
-kz_2+F^{\textrm B}_2
\end{bmatrix}
/{(6\pi\eta_\textrm s ag_0)}.
\end{eqnarray}
The eigenvalues of the mobility tensor are $1/6\pi\eta_\textrm s
ag^\textrm I_\|$, $1/6\pi\eta_\textrm s a g_\|^\textrm{II}$,
$1/6\pi\eta_\textrm s a g_\bot^\textrm{I}$, and $1/6\pi\eta_\textrm
s a g_\bot^\textrm{II}$, which represent the inverse hydrodynamic
resistance in the four hydrodynamic modes with boundary conditions
described in Eq. \ref{E-BC}. The four principal vectors are
$Y_\|^\textrm I=x_1+x_2$, $Y_\|^\textrm {II}=x_1-x_2$,
$Y_\bot^\textrm I=z_1+z_2$, and $Y_\bot^\textrm{II}=z_1-z_2$,
representing the common (collective, mode I) and relative (mode II)
motions of spheres in both parallel and perpendicular directions,
respectively. The eigenvector lead to the following autocorrelation
functions for the pair interactions under the four eigenmodes:
\begin{eqnarray}
&&\displaystyle{\langle Y_\|^\textrm I(t)Y_\|^\textrm I(0)\rangle =
\frac{2k_BT}{k}\left[\exp\left(-\frac{t}{\tau_\|^\textrm I}
\right)\right]},\\
&&\displaystyle{\langle
Y_\|^\textrm{II}(t)Y_\|^\textrm{II}(0)\rangle =
\left(\frac{2{F^\textrm{dep}}}{k}\right)^2 +
\frac{2k_BT}{k}\left[\exp\left(-\frac{t}{\tau_\|^\textrm {II}}
\right)\right],}\nonumber\\
&&\displaystyle{\langle Y_\bot^\textrm I(t)Y_\bot^\textrm
I(0)\rangle =
\frac{2k_BT}{k}\left[\exp\left(-\frac{t}{\tau_\bot^\textrm I}
\right)\right],\quad\textrm{and}}\nonumber\\
&&\displaystyle{\langle Y_\bot^\textrm {II}(t)Y_\bot^\textrm
{II}(0)\rangle =
\frac{2k_BT}{k}\left[\exp\left(-\frac{t}{\tau_\bot^\textrm {II}}
\right)\right],\nonumber}
\end{eqnarray}
where the individual decay time for each mode are $\tau_\|^\textrm I
=6\pi\eta_\textrm s g^\textrm{I}_{\|}/k$, $\tau_\|^\textrm {II}
=6\pi\eta_\textrm s g^\textrm{II}_{\|}/k$, $\tau_\bot^\textrm I
=6\pi\eta_\textrm s g^\textrm{I}_{\bot}/k$, and $\tau_\bot^\textrm
{II} =6\pi\eta_\textrm s g^\textrm{II}_{\bot}/k$. The four decay
times of the corresponding principal modes all reduce to the single
particle limit as the separation distance becomes large, e.g., $r/a>
\mathcal O(10^2)$. Figure 7 demonstrates autocorrelation functions
under the four principal modes with various concentration and
separation distance. Comparing Figs 7a with 7b, and 7c with 7d, in
both directions the decorrelation times in mode I (the collective
motion, 7a and 7c) are always shorter than mode II (relative motion,
7b and 7d). The reason is because the hydrodynamic resistance is
always higher in the relative motion, which slows down the particle
motion more significantly especially in the lubrication regime.
\begin{figure}[ht]
\begin{center}
\includegraphics[width=6.4in]{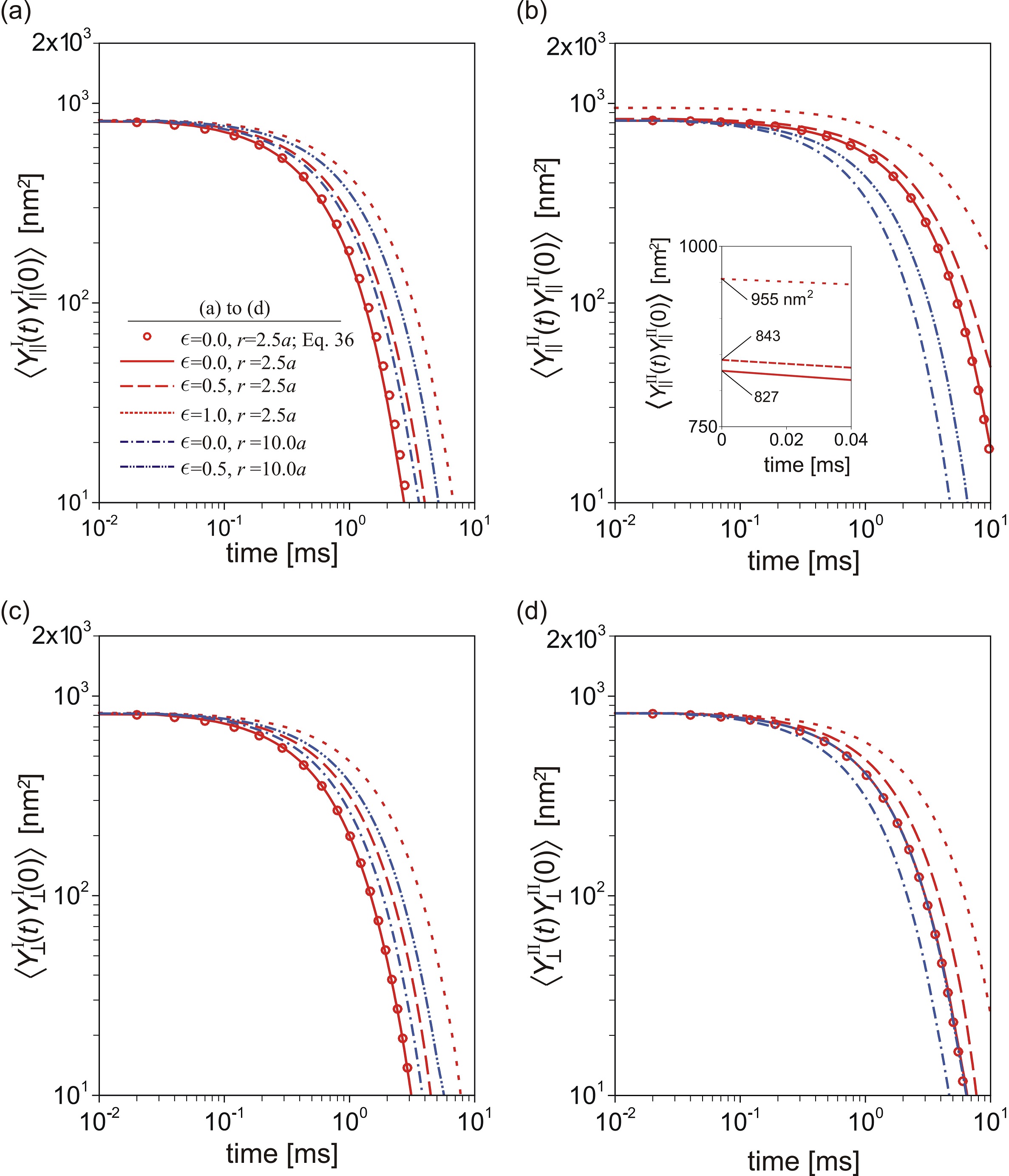}
\end{center}
\caption{Autocorrelation functions of the displacement along the
four principal directions. The data points represent the analytical
results (Eq. 29) using the mobility functions given by
Batchelor~\cite{batchelor1976brownian}. Parameters for particles and
polymer are the same as in Fig. \ref{F-single-auto}.}
\label{F-corr-nor}
\end{figure}
When comparing parallel to perpendicular motions, in mode I the
behavior is similar for both motions because the depletion effect
has been canceled in both eigenmodes $Y_\|^\textrm{I}$ and
$Y_\|^\textrm{II}$ and therefore only the hydrodynamic effect plays
a role in mode I, in which the correction factors in parallel and
perpendicular motions have similar values. In the eigenmode II
parallel motion (7b), the autocorrelation function includes the
decomposed contributions from two sources, the depletion force and
the hydrodynamic resistance. The polymer depletion effect changes
the equilibrium position of spheres along $Y_\|^\textrm{II}$
direction with a magnitude of $2{F^\textrm{dep}}/k$. This shifts the
autocorrelation of spheres' movement in parallel mode II with a
magnitude of $4({F^\textrm{dep}}/k)^2$. In nearby location of
optical traps ($r/a\simeq2.5$) the decorrelation time of motion of
spheres is enhanced. In mode II, the increase of the decorrelation
time is more pronounced in parallel (7b) than in the perpendicular
(7d) directions. It is due to the larger hydrodynamic correction
factor $g_\|^\textrm{II}$ compared to $g_\bot^\textrm{II}$, which
can be speculated from the $g$-factor plots (Figs. \ref{F-Hyd-int}b,
\ref{F-Hyd-int}c). With increasing polymer concentration $\epsilon$
the decorrelation time in all modes will increase due to larger
viscous resistance force. In principle one may compute the $g$
factor and the decay time of displacement autocorrelation functions
in four principal directions by experimentally measuring the auto-
and cross-correlation of displacements. It is also valuable to
validate the attractive potential between two hard spheres from the
dynamic view point through the shift $4({F^\textrm{dep}}/k)^2$ in
the autocorrelation function ($Y_\|^\textrm{II}$).

\begin{figure}[ht]
\begin{center}
\includegraphics[width=6.4in]{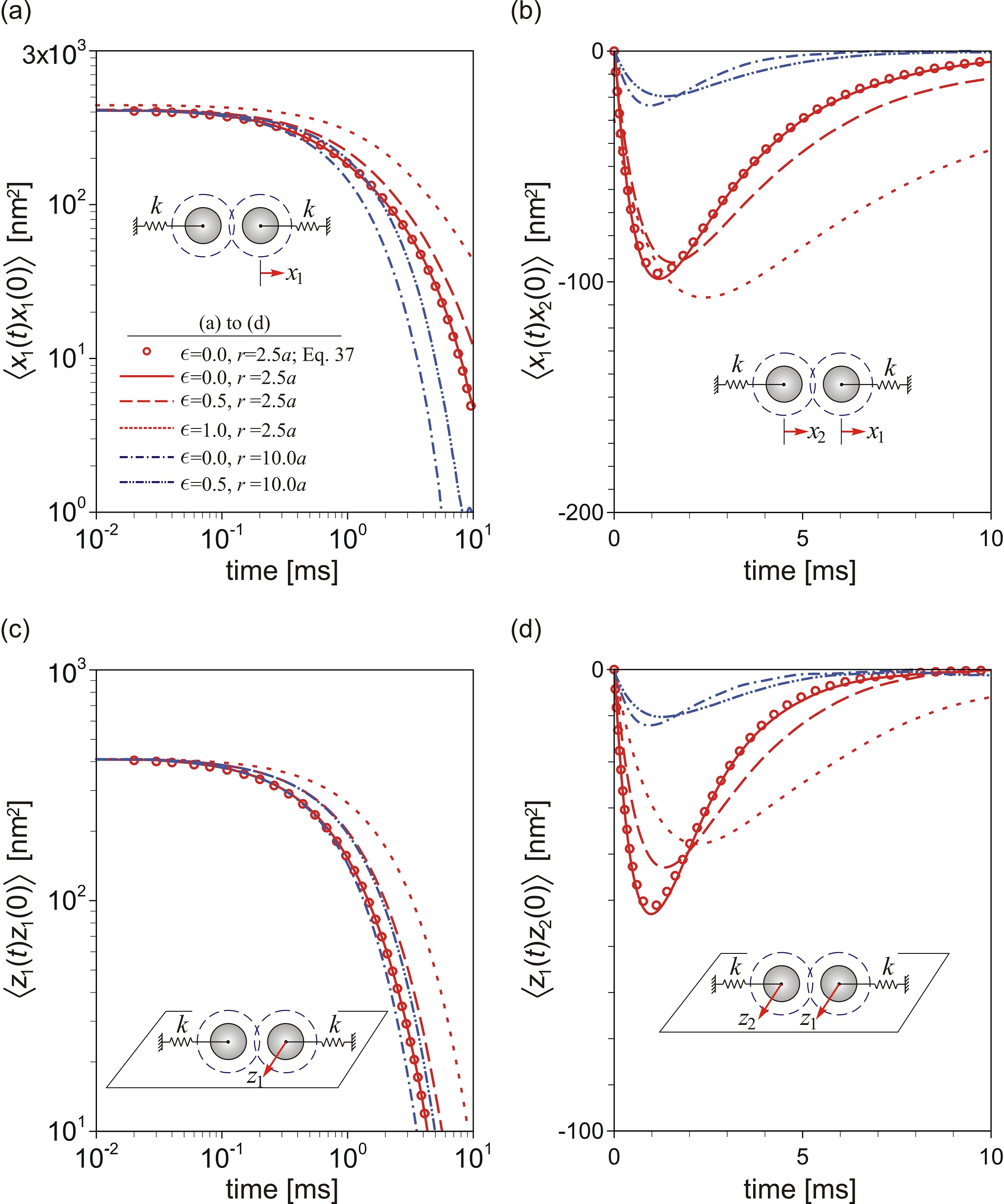}
\end{center}
\caption{Correlation functions of the displacement of two trapped
spheres in the $xz-$plane under the influence of depletion layers.
Parameters used for particles and polymer conditions are the same as
in Fig. \ref{F-single-auto}.} \label{F-paircorr}
\end{figure}

In physical space, the correlation functions on the $x,z$ plane are
\begin{eqnarray}
&&\displaystyle{\langle x_1(t)x_1(0)\rangle =\left(
\frac{{F^\textrm{dep}}}{k}\right)^2+\frac{k_BT}{2k}\left[\exp\left(-\frac{t}{\tau_\|^\textrm{I}}
\right) + \exp\left(-\frac{t}{\tau_\|^\textrm{II}}\right)\right],}\\
&&\displaystyle{\langle x_1(t)x_2(0)\rangle =
-\left(\frac{{F^\textrm{dep}}}{k}\right)^2+\frac{k_BT}{2k}\left[\exp\left(-\frac{t}{\tau_\|^\textrm{I}}\right)
-\exp\left(-\frac{t}{\tau_\|^\textrm{II}}\right)\right],}\nonumber\\
&&\displaystyle{\langle z_1(t)z_1(0)\rangle =
\frac{k_BT}{2k}\left[\exp\left(-\frac{t}{\tau_\bot^\textrm{I}}\right)
+\exp\left(-\frac{t}{\tau_\bot^\textrm{II}}\right)\right],}\quad\textrm{and}\nonumber\\
&&\displaystyle{\langle z_1(t)z_2(0)\rangle =
\frac{k_BT}{2k}\left[\exp\left(-\frac{t}{\tau_\bot^\textrm{I}}\right)
-\exp\left(-\frac{t}{\tau_\bot^\textrm{II}}\right)\right].}\nonumber
\end{eqnarray}
The correlation results are plotted in Fig \ref{F-paircorr}. Note
correlation functions above include the self- and cross-correlations
in the principle directions, and can be computed by
\begin{eqnarray}
\displaystyle{\langle x_1(t)x_1(0)\rangle = \frac{1}{4}\left[\langle
Y_\|^\textrm I(t)Y_\|^\textrm I(0)\rangle+\langle Y_\|^\textrm
I(t)Y_\|^\textrm {II}(0)\rangle+\langle Y_\|^\textrm
I(0)Y_\|^\textrm {II}(t)\rangle +\langle Y_\|^\textrm
{II}(t)Y_\|^\textrm {II}(0)\rangle\right]},
\end{eqnarray}
and \begin{eqnarray} \displaystyle{\langle x_1(t)x_2(0)\rangle =
\frac{1}{4}\left[\langle Y_\|^\textrm I(t)Y_\|^\textrm
I(0)\rangle-\langle Y_\|^\textrm I(t)Y_\|^\textrm
{II}(0)\rangle+\langle Y_\|^\textrm I(0)Y_\|^\textrm {II}(t)\rangle
-\langle Y_\|^\textrm {II}(t)Y_\|^\textrm {II}(0)\rangle\right]}.
\end{eqnarray}\label{E-cor-lab}
\noindent The same formulations are applied to find the correlations
functions in the $z$-direction. Along the parallel direction (Fig.
8a and 8b) the auto- and cross-correlation functions at $t=0$
deviates from $k_\textrm B T/k$ (auto) and zero (cross) with an
amount of $({F^\textrm{dep}}/k)^2$, however, the two decorrelation
time scales involved maybe difficult to distinguish from each other.
Similarly in the perpendicular direction, the correlation functions
are determined by two time scales that originate from the two
hydrodynamic modes. Considering the decayed hydrodynamic
interactions, at very long lag time all correlations vanish, while
at zero lag time the cross-correlation vanishes, implying that
apparently the particle does not feel the presence of another
particle at the beginning due to the mutual cancellation of
collective and relative motions. The entropic force has a finite
influence on both correlations. They are always anti-correlated
because the first mode (collective motion) always decays faster than
the second mode (relative motion) due to its less resistance. In
other words, this relative motion dominates the relevance of the
displacements of both particles after certain lag time, and the pair
interaction behaves like in its second or anticorrelated mode. This
fact is not to be confused with the intuition obtained from a steady
mobility analysis.

The largest anticorrelation appears at
\begin{equation}\label{E-cross-m-t}
t_{\textrm{min},\|} = \frac{\tau^\textrm{I}_\|\tau^\textrm{II}_\|}
{\tau^\textrm{I}_\|- \tau^\textrm{II}_\|}\ln
\left(\frac{\tau^\textrm{I}_\|}{\tau^\textrm{II}_\|}\right) =
\frac{g^\textrm{I}_\|g^\textrm{II}_\|} {g^\textrm{I}_\|-
g^\textrm{II}_\|}\ln
\left(\frac{g^\textrm{I}_\|}{g^\textrm{II}_\|}\right).
\end{equation}
The same formulation is applicable for the perpendicular direction.
The depth of the anticorrelation in parallel direction versus the
distance of optical traps is predicted in Fig. 9. At large
separation distance the motion obviously is uncorrelated. At short
separation distance, the increase of the depth is due to both
hydrodynamic and depletion interactions. Without the depletion
effect, the analytical results from Batchelor's approximation (solid
curves) is somewhat higher than the numerical simulation (black
circles) because the approximation yields a relative lower
lubrication force, which enhances the influence of the relative
motion. Consistent results are shown by the two blue lines in Fig.
8b and 8d, in which the higher viscosity gives a shallower
anticorrelation. By bringing the pair particles closer to each
other, the minimum value for $\langle x_1(t)x_2(0)\rangle$ is lower
at small $a/r$ and then higher for the polymer solution cases with
depletion effect (red squares and blue diamonds) compared with the
solvent-only case (black circles). The overlapping of the depletion
zones appears at $a/r$$\approx$0.31 ($\epsilon=0.5$) and
$a/r$$\approx$0.38 ($\epsilon=1.0$). Without depletion zones, the
uniform polymer solution case, the hydrodynamic resistance
experience by the spheres is higher because the higher bulk
viscosity simply gives a shallower anticorrelation. Into the overlap
region, the entropic force brings particles toward each other and
enhances anticorrelation mode. This effect is further enhanced by
bringing the spheres closer. The complicated flow pattern and
resistance in the two-layer model has relatively much less
contribution compared with the entropic effect on the
anticorrelation depth. Experimental validation is needed to validate
this result.

\begin{figure}[ht]
\begin{center}
\includegraphics[width=3.2in]{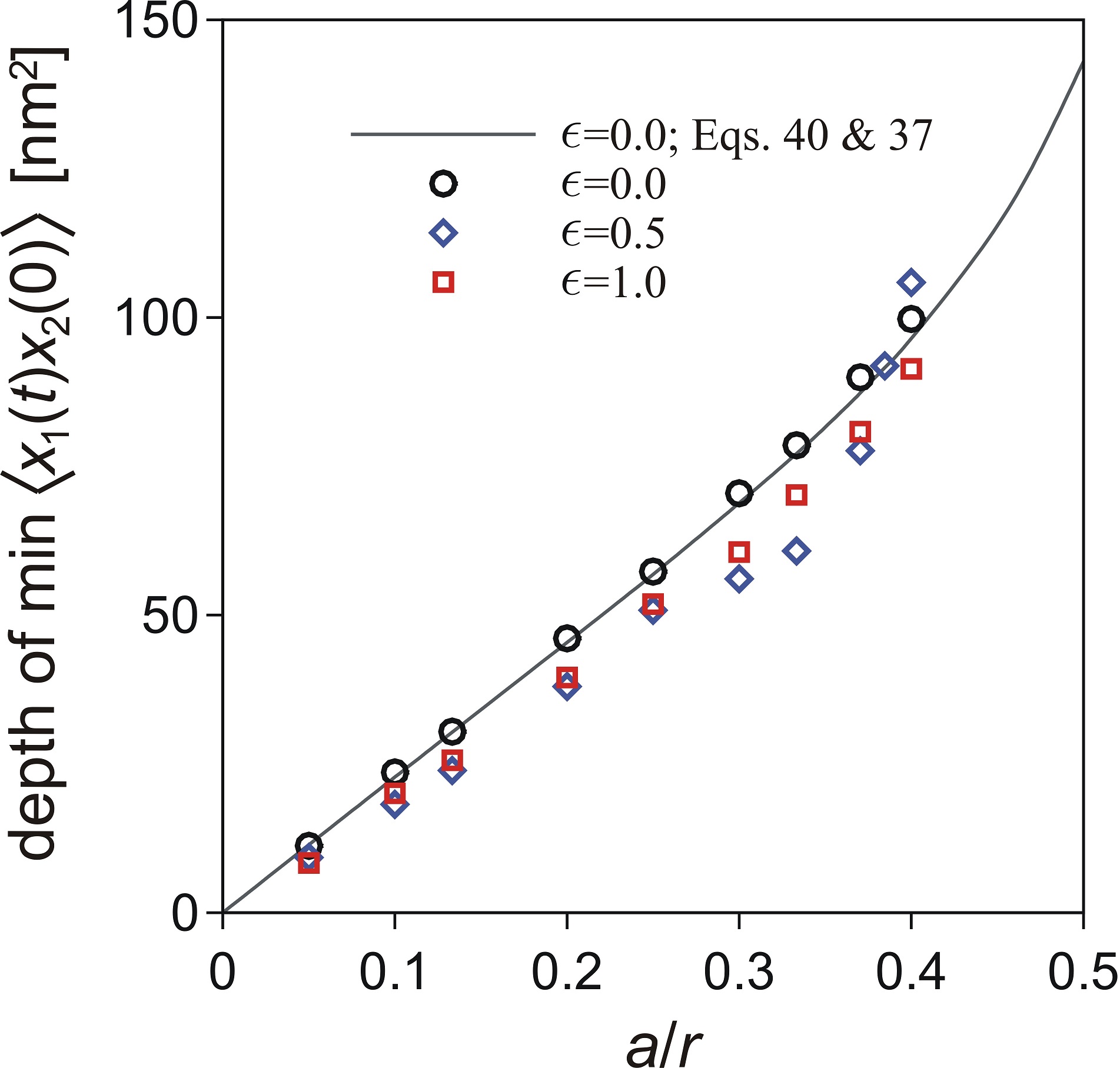}
\end{center}
\caption{The maximum value of the anticorrelation in the parallel
direction versus the center-to-center separation distance. The limit
$a/r=0$ indicates an isolated particle, while $a/r=0.5$ means both
particles are in contact with each other. The reference curve is the
analytical approximation based on the mobility function for the
solvent-only case.} \label{F-mincro}
\end{figure}

\section{Conclusion}
We presented a theoretical description that enables to characterize
the stochastic motions of a pair of hard spheres in non-adsorbing
polymer solutions. Based on a two-layer  approximation for the
polymer depletion effect, the hydrodynamic mobility tensor is
resolved by the boundary integral analysis, which determines the
corresponding time evolution of stochastic displacements that are
consistent with the fluctuation-dissipation theorem. We first find
that the presence of depletion zones significantly modifies the
hydrodynamic interactions between two moving spheres. Polymer
depletion is found to shift the probability distribution of the
random walks of two trapped spheres. The resulting auto- and
cross-correlation functions are presented in the principal modes,
which clearly identify the decomposed entropic and hydrodynamic
effects on the dynamic behavior of two hard spheres under the
influence of added polymers to the solvent for various particle
separation distances. In the presence of depletants, the
cross-correlation of the particle displacements is weakened in a way
that higher viscosity slows down the motion significantly and
lengthens the hydrodynamic decay time. However, the appearance of
the depletion zone reduces this viscous effect. Finally, the
cross-correlation is significantly enhanced when the depletion zones
overlap.\\

\noindent\textbf{Acknowledgments}~~M. Karzar-Jeddi and T.-H. Fan
acknowledge the support from NSF CMMI-0952646, R. Tuinier thanks DSM
for support, and T. Taniguchi acknowledges the support from JSF.

\newpage
\begin{appendices}
\section*{APPENDIX A}
\renewcommand{\theequation}{A\arabic{equation}}
\setcounter{equation}{0}

The divergence of the diffusivity tensor (Eqs. \ref{E-diff-ten-1}
and \ref{E-diff-ten-2}) can be expressed as
\begin{eqnarray}\label{E-der1}
\frac{\partial}{\partial \mathbf{r}_\alpha}\cdot\mathbf
D_{\alpha\alpha}=&& \frac{D_0}{g_0} \left.\Bigg[ \frac{\mathbf
r_{\alpha\beta}\mathbf r_{\alpha\beta}}{r^2}\cdot\frac{\partial
A^\textrm s}{\partial\mathbf r_{\alpha}} +\left(\mathbf I
-\frac{\mathbf r_{\alpha\beta}\mathbf
r_{\alpha\beta}}{r^2}\right)\cdot\frac{\partial B^\textrm
s}{\partial\mathbf r_{\alpha}}\right. \\
&&~~+\left. (A^\textrm s-B^\textrm s)\mathbf r_{\alpha\beta}\mathbf
r_{\alpha\beta}\cdot\frac{\partial~~}{\partial \mathbf
r_\alpha}\left(\frac{1}{r^2}\right)+\left(\frac{A^\textrm
s-B^\textrm s}{r^2}\right)\frac{\partial~~}{\partial\mathbf
r_\alpha}\cdot(\mathbf r_{\alpha\beta} \mathbf
r_{\alpha\beta})\right. \Bigg]\nonumber,
\end{eqnarray}
where $\alpha,\beta=1,2$ and $\alpha\neq\beta$, $\mathbf
r_{\alpha\beta}=\mathbf r_\beta-\mathbf r_\alpha$, and $r=|\mathbf
r_\beta-\mathbf r_\alpha|$. By simplifying the differential terms on
the right
\begin{eqnarray}\label{E-der2}
\frac{\partial}{\partial \mathbf r_{\alpha}}\cdot\mathbf
D_{\alpha\alpha}&=&\frac{D_0}{g_0} \left.\Bigg\{ \frac{\mathbf
r_{\alpha\beta}\mathbf
r_{\alpha\beta}}{r^2}\cdot\left(\frac{-\mathbf
r_{\alpha\beta}}{r}\frac{\partial A^\textrm s}{\partial r}\right)
+\left(\mathbf I -\frac{\mathbf r_{\alpha\beta}\mathbf
r_{\alpha\beta}}{r^2}\right)\cdot\left(\frac{-\mathbf
r_{\alpha\beta}}{r}\frac{\partial B^\textrm s}{\partial
r}\right)\right. \\
&&~~~~+\left. (A^\textrm s-B^\textrm s)\mathbf
r_{\alpha\beta}\mathbf r_{\alpha\beta}\cdot\left(\frac{2\mathbf
r_{\alpha\beta}}{r^4}\right)+\frac{A^\textrm s-B^\textrm
s}{r^2}\Bigg[-\mathbf r_{\alpha\beta}\cdot\mathbf I +\mathbf
r_{\alpha\beta}\left(\frac{\partial~~}{\partial\mathbf
r_\alpha}\cdot\mathbf
r_{\alpha\beta}\right)\Bigg]\right.\Bigg\}\nonumber\\
\nonumber\\&=&\frac{D_0}{g_0} \Bigg[\frac{-r^2\mathbf
r_{\alpha\beta}}{r^3}\frac{\partial A^\textrm s}{\partial r}
+\frac{-\mathbf r_{\alpha\beta}}{r}\frac{\partial B^\textrm
s}{\partial r}+\frac{r^2\mathbf r_{\alpha\beta}}{r^3}\frac{\partial
B^\textrm s}{\partial r}+\frac{2(A^\textrm s-B^\textrm
s)}{r^2}\mathbf r_{\alpha\beta}+\frac{-4(A^\textrm s-B^\textrm
s)}{r^2}\mathbf r_{\alpha\beta}\Bigg].\nonumber
\end{eqnarray}
Therefore
\begin{eqnarray}
\frac{\partial}{\partial \mathbf r_{\alpha}}\cdot\mathbf
D_{\alpha\alpha}&=& \frac{-D_0}{g_0}\left[ \frac{\partial A^\textrm
{s}}{\partial r}+\frac{2(A^\textrm s-B^\textrm
s)}{r}\right]\frac{\mathbf r_{\alpha\beta}}{r}.
\end{eqnarray}
Similarly, the divergence of the mutual diffusivity tensor can be
expressed as
\begin{eqnarray}\label{E-der2}
\frac{\partial}{\partial \mathbf r_\beta}\cdot\mathbf
D_{\alpha\beta}&=&\frac{D_0}{g_0} \Bigg\{ \frac{\mathbf
r_{\alpha\beta}\mathbf r_{\alpha\beta}}{r^2}\cdot\left(\frac{\mathbf
r_{\alpha\beta}}{r}\frac{\partial A^\textrm c}{\partial r}\right)
+\left(\mathbf I -\frac{\mathbf r_{\alpha\beta}\mathbf
r_{\alpha\beta}}{r^2}\right)\cdot\left(\frac{\mathbf
r_{\alpha\beta}}{r}\frac{\partial B^\textrm c}{\partial
r}\right) \\
&&~~~~+(A^\textrm c-B^\textrm c)\mathbf r_{\alpha\beta}\mathbf
r_{\alpha\beta}\cdot\left(\frac{-2\mathbf
r_{\alpha\beta}}{r^4}\right)+\frac{A^\textrm c-B^\textrm
c}{r^2}\Bigg[\mathbf r_{\alpha\beta}\cdot\mathbf I +\mathbf
r_{\alpha\beta}\left(\frac{\partial~~}{\partial\mathbf
r_\beta}\cdot\mathbf r_{\alpha\beta}\right)\Bigg]\Bigg\}.\nonumber
\end{eqnarray}
Which simplifies to Eq. \ref{E-graddiff-2},
\begin{eqnarray}
\frac{\partial}{\partial \mathbf r_{\beta}}\cdot\mathbf
D_{\alpha\beta}&=& \frac{D_0}{g_0}\left[ \frac{\partial A^\textrm
{c}}{\partial r}+\frac{2(A^\textrm c-B^\textrm
c)}{r}\right]\frac{\mathbf r_{\alpha\beta}}{r}.
\end{eqnarray}
In a uniform fluid ($\lambda$=1), $A^\textrm s$=$B^\textrm s$=0,
$A^\textrm c=3a/2r$, $B^\textrm c=3a/4r$ for the Oseen tensor, and
$A^\textrm s$=$B^\textrm s$=0, $A^\textrm c=3a/2r-a^3/r^3$,
$B^\textrm c=3a/4r+a^3/2r^3$ for Rotne-Prager tensor. For both cases
$\partial \mathbf D_{\alpha\alpha}/\partial\mathbf r_\alpha =
\partial \mathbf D_{\alpha\beta}/\partial\mathbf r_\beta=0$ are applicable in Brownian motion simulations..
\end{appendices}

\bibliographystyle{unsrt}

\end{document}